\documentclass[aps,prd,groupedaddress,onecolumn,showpacs,10pt]{revtex4-1}

\usepackage{graphicx,color}
\usepackage{bm}
\usepackage{amssymb,amsmath,amsfonts, mathrsfs}

\newcommand{\be}{\begin{equation}}
\newcommand{\ee}{\end{equation}}
\newcommand{\ba}{\begin{array}}
\newcommand{\ea}{\end{array}}
\newcommand{\bqa}{\begin{eqnarray}}
\newcommand{\eqa}{\end{eqnarray}}

\begin{document}

\begin{flushright}
USTC-ICTS/PCFT-20-24\\
\end{flushright}

\title{Relativistic Friedrichs-Lee model and quark-pair
creation model}


\author{Zhi-Yong Zhou}
\email[]{zhouzhy@seu.edu.cn}
\affiliation{School of Physics, Southeast University, Nanjing 211189,
P.~R.~China}
\author{Zhiguang Xiao\footnote{Corresponding author}}
\email[]{xiaozg@ustc.edu.cn}
\affiliation{NSFC-SFTP Peng Huanwu Center for Fundamental Theory and Interdisciplinary Center for Theoretical Study, University of Science
and Technology of China, Hefei, Anhui 230026, China}


\date{\today}

\begin{abstract}
In this paper, we present how the Friedrichs-Lee model could be
extended to the relativistic scenario and be combined with the
relativistic quark pair creation model in a consistent way. This
scheme could be applied to study the ``unquenched'' effect of the meson
spectra.  As an example, if the lowest $J^{PC}=0^{++}$ $(u\bar u+d\bar
d)/\sqrt{2}$ bound state in the potential model is coupled to the
$\pi\pi$ continuum, two resonance poles could be found from the
scattering amplitude for the continuum states. One of them could
correspond to the $f_0(500)/\sigma$ and the other probably
$f_0(1370)$.  This scheme might shed more light on why extra states
could appear in the hadron spectrum other than the prediction of the
quark potential model.
\end{abstract}


\maketitle
\section{Introduction}
The quark potential models, by introducing the interactions respecting
the properties of the quantum chromodynamics~(QCD), have achieved a
general success in predicting many meson states with different quantum
numbers~\cite{Eichten:1978tg,Godfrey:1985xj}. Especially, taking into
account the relativistic effect, the Godfrey-Isgur~(GI) model
provides a unified description of most of mesons and presents  fairly
reasonable predictions to their masses. The severe deviations from the
experimental observation only happen in two regions: the first one is
the long-history puzzle of identifying the light scalar sectors; the
second one is about the new exotic charmonium-like states starting
from the observation of $X(3872)$, especially for the states above the
open-flavor thresholds. We are aiming at the former region in this
paper, while part of the latter one has been addressed in some other
works~\cite{Zhou:2017dwj,Zhou:2017txt}.

The lowest scalar meson predicted in the potential models is located
at above 1.0 GeV~\cite{Godfrey:1985xj}. However, the $I=0$ $\pi\pi$
scattering phase shift rises smoothly and passes $90^\circ$ at about
850 MeV~\cite{Protopopescu:1973sh,Grayer:1974cr,Becker:1978ks}, { so it
was believed that there exists a broad structure contributing to
the phase shift in the
energy region from the $\pi\pi$ threshold to above 1.0 GeV}~\cite{Minkowski:1998mf}. Many
phenomenological studies have been devoted to proving the existence of
this structure which is dubbed $f_0(500)$ now in the
particle data group~(PDG) table~\cite{Tanabashi:2018oca}, and its pole position, is confirmed
and determined by model-independent methods such as in
Refs.\cite{Zhou:2004ms,Caprini:2005zr}. The existence of $f_0(500)$
and $f_0(980)$ with clear experimental evidences made people lose
faith in the  predictions of light scalar mesons in the potential
model. Thus, people usually believe that identifications of the light
scalar mesons are totally a mess plagued by strong overlap between
resonance and background,  and the quark model does not work here
completely.
Moreover, the $I=1/2$ $K\pi$ phase shifts measured from about 100 MeV
above the threshold in the $Kp$ production also exhibit a similar
smoothly-rising behavior~\cite{Estabrooks:1977xe}, which leads to the
discovery of another board structure, denoted as $K_0^*(700)$ or $\kappa$
now, whose pole position is also determined more and more
accurately~\cite{Zheng:2003rw,DescotesGenon:2006uk,Pelaez:2020uiw}.

In order to understand such states, several different kinds
of methods were introduced. The tetraquark model, proposed by
Jaffe~\cite{Jaffe:1976ig}, regarding them as fourquark states produced
by QCD fundamental interaction, was adopt to understand their masses, and
$f_0(500)/\sigma$, $K_0^*(700)/\kappa$, $a_0(980)$, and  $f_0(980)$
are regarded as lightest tetraquark nonet~\cite{Maiani:2004uc}.
Another idea is that the pseudoscalar-pseudoscalar meson scattering
could be well described by the chiral perturbation
theory~($\chi$PT)~\cite{Gasser:1983yg}, and the resonance information
could be restored by unitarizing the $\chi$PT amplitudes with some
unitarization schemes~\cite{Oller:1998hw,Guo:2011pa}. In this picture,
the $\sigma$, $\kappa$ resonances can be viewed as dynamically
generated from the $\pi\pi$ or $\pi K$ interaction rather than the
fundamental states produced by QCD.  However, this is in the picture of
the effective interaction of Goldstones rather than from a constituent
quark point of view.
Although this kind of method has provided plausible explanations of
these resonant states, one still wonders whether these states can
really be dynamically generated by the interactions of
the meson states consistently from the point of view of the
constituent quark model which
captures the nature of the mesons with various quantum numbers in a
much broader range.
In fact, in GI's paper, they already noticed that the spectra
produced by the potential model did not included the interactions
between the mesons and the nearby continuum which may modify the mass
spectra and produce the width for the mesons.
From recent experience in studying the exotic heavy quarkonium-like states
like $X(3872)$, it is also demonstrated that by coupling the QCD fundamental
$q\bar q$ states with the continuum, not only can the fundamental state
itself be modified from the potential model predictions, but also new
states can be dynamically generated. This idea was successfully used in
explaining the generation of the $X(3872)$~\cite{Kalashnikova:2005ui,Ortega:2009hj,Takizawa:2012hy,Coito:2012vf,Sekihara:2014kya,
Zhou:2017dwj,Zhou:2017txt,Giacosa:2019zxw}.
Actually, this kind of idea has already been widely used in studying the charmonium state~\cite{Eichten:1978tg} and the
low lying scalar mesons~\cite{vanBeveren:1982qb, vanBeveren:1986ea,
Tornqvist:1995kr,Zhou:2010ra} with some successes.
In fact, even long before these practices, the fact that coupling the discrete
state to the continuum will change the spectrum was generalized and
demonstrated in the so-called Friedrichs-Lee model~\cite{Friedrichs:1948,Lee:1954iq}.

In 1948, Friedrichs established the simplest form of the model in a
non-relativistic scenario~\cite{Friedrichs:1948}, in which the free
Hamiltonian has a discrete eigenstate and a continuum eigenstate, and
an interaction between the discrete state and the continuum states is
introduced. The eigenvalue of the discrete state is embedded in the
continuous mass spectrum of the continuum when the interaction is
turned off. Once the discrete state and
the continuum state are coupled with each other, the discrete state
will dissolve into the continuum state and becomes an unstable state
with a certain width. This model is exactly solvable so that some
properties of unstable states, such as the wave function of the
resonance or scattering amplitudes for the continuum states, could be
studied carefully. One of the most important ingredients of the model
is the resolvent function $\frac{1}{\eta(E)}$ as a function of the energy
$E$, with
\bqa
\eta(E)=E-\omega_0-\int \mathrm{d}E\frac{\rho(E')}{E-E'}.
\label{eq:nonrela-resolvent}
\eqa
When the function is analytically continued to the complex $E$ plane,
zero points of $\eta(E)$ function represent the poles of the
scattering amplitude of the continuum states. When the bare discrete
state is below the threshold, there is also a virtual state accompanied
 with the original discrete state as the
interaction is turned on. The importance of the extra
virtual-state, bound-state, and resonance poles, which could appear in
the amplitude besides the one originally at $\omega_0$, has been
emphasized in Ref.\cite{Xiao:2016mon,Xiao:2016wbs,Xiao:2016dsx}. These
so-called  Gamow states, denoted by their complex pole positions on the
unphysical Riemann sheets, are generalized eigenstates of the
Hamiltonian, which has a good definition in the rigged
Hilbert space~\cite{Civitarese200441,Bohm:1989}.

Similar excellent ideas were proposed independently by several theorists
in different areas in physics. In the quantum field theory, the Lee
model is established to study how the processes depicted by
$V\rightleftarrows N+\theta$ could influence the physical state and
wave function renormalization~\cite{Lee:1954iq}. The Feshbach
resonance theory~\cite{Feshbach:1958nx} and the Anderson
model~\cite{Anderson:1961} are also different forms with similar
spirits in nuclear physics and condensed matter physics. In hadron
physics, similar non-relativistic methods originated from this idea
have been applied or developed by several different
groups~\cite{Kalashnikova:2005ui,Ortega:2009hj,Takizawa:2012hy,Sekihara:2014kya,Zhou:2017dwj,Xiao:2016mon,
Wolkanowski:2015jtc,Wolkanowski:2015lsa},
in understanding some charmonium-like state, especially for the
enigmatic $X(3872)$ state.

To utilize this Friedrichs-Lee scheme in the low lying mesons states,
the relativistic effects need to be considered.
This is because in the light meson states with $u$, $d$, and $s$
quarks, the constituents are light and might travel very fast, so that the
non-relativistic methods might not be self-consistent. There are two
aspects of relativistic effects to be considered. The first is
related to the Friedrichs-Lee model itself, in particular,
 the resolvent function. It is well known that the
relativistic dispersion relation is expressed in $s$,
the invariant total momentum squared,  and thus
the corresponding resolvent function should be
expressed in $s$ rather than $E$ as in (\ref{eq:nonrela-resolvent}).
This is due to the presence of antiparticles or annihilation operators
in relativistic theory.
There are several ways to incorporate the relativistic effects into the
Friedrichs model~\cite{Horwitz:1994bs,Antoniou:1998JMP} in a more
systematical way and we will
follow the method in \cite{Antoniou:1998JMP} which is most direct and
simple by introducing a bilocal operator to simulate the
two-particle state. The second aspect is related to the interaction between
the discrete state and the continuum, which should be modeled
 with some relativistic effects taken into account. The well-known nonrelativistic quark pair creation~(QPC) model which
was used in the discussion of the heavy charmonium states can not be
directly used here and should be modified to take into account some
relativistic effects.
T\"ornqvist
developed a unitarized quark model which introduce a meson propagator
with a relativistic dispersion
relation~\cite{Tornqvist:1995kr,Heikkila:1983wd}
\bqa
P(s)^{-1}=s-m_A^2-\int_{s_{th}} \mathrm{d}s\frac{f^{A}_{BC}(s')}{s-s'}
\eqa
to describe how the full propagator of meson $A$ be influenced by
coupling to the meson pair $BC$. $m_A$ is the bare mass of meson $A$,
$s_{th}$ is the energy squared of the $BC$ threshold, and
$f^{A}_{BC}(s)$ is the spectral function determined by the $A-BC$
coupling form factor. Although  the dispersion
relation is a relativistic form, $f^{A}_{BC}$ is derived from the non-relativistic QPC
interaction. In ref.~\cite{vanBeveren:1982qb}, Beveren et. al. took into
account the Lorentz transform of the energy in the wavefunction in
the QPC model. A more
thorough and complete treatment of the relativistic effect on the states
and wavefunction in the QPC model was carried out by Fuda, in which the Lorentz
transformations are taken into account when the constituent quarks and
antiquarks regroup to form a new meson pair~\cite{Fuda:2012xd}.
However, his trial study of the $\rho$
meson coupling to $\pi\pi$, a non-relativistic dispersion relation was
adopt.

In our paper, we will combine Fuda's relativistic QPC approach and the relativisitic
generalization of the Friedrichs-Lee model.
This relativistic Friedrichs-Lee-QPC scheme is then
applied to study the lowest $I=0$, $J^{PC}=0^{++}$, $(u\bar u+d\bar d)/\sqrt{2}$
bound state, with the GI model as the input, coupling to the $\pi\pi$
continuum state, and it is found that while the discrete state is
shifted onto
the complex $s$-plane, a light broad resonance pole corresponding
to $f0(500)/\sigma$ could be generated naturally in the $\pi\pi$ scattering
amplitude. Furthermore, it is observed that the two poles contribute a mild total
phase shift of about $180^\circ$.  Their contributions to the smooth rise of
the phase shift is consistent with the one measured in the $\pi p$
experiments~\cite{Protopopescu:1973sh,Grayer:1974cr,Becker:1978ks}.

The paper is organized as follows: The theoretical background is
briefly introduced in Section \ref{theory}. To prepare for the
relativistic treatment of the Friedrichs-Lee model and QPC model, the
definitions of relativistic canonical single-particle and two-particle
states are presented in Section \ref{statesdef}. Then the relativistic
Friedrichs-Lee model is introduced and solved in Section
\ref{relativisticFM}. The readers could just skip the details of the
formal deduction of the solution and jump to the conclusion at the end
of this section if they wish. The relativistic QPC model is briefly
reviewed in Section \ref{relqpc} and the coupling form factor is
obtained in this scheme. The numerical calculation and its application
in studying the $f_0(500)$ state are briefly discussed in Section
\ref{discussion}. Some Lorentz transformation and kinematics used in
the calculation are described in detail in
the appendix \ref{appendixB}.

\section{Theoretical background}\label{theory}
Since both the Friedrichs-Lee model and the QPC model need to deal with
relativistic two-particle states, either at the hadron level or at the
quark level,  we first introduce the definitions of canonical
one-particle and two-particle states and their transformation
properties under the Lorentz transformation used in this paper. We
then present the relativistic Friedrichs-Lee model in the angular
momentum representation, and finally review the relativistic quark
pair creation~(QPC) model, which is used to describe the interaction
in the relativistic Friedrichs-Lee model. This section is written in a
pedagogical manner to make the presentation self-consistent and easily
understandable.
\subsection{Definition of one-particle and two-particle states}
\label{statesdef}
The one-particle and two-particle states in this paper are represented
by the canonical states but not by the helicity states. This means that the
third-component of the spin of the state is defined along a fixed direction, the
$z$-axis, in the rest frame of the particle. Such a choice is for the
sake of the
convenience in  the discussion of  the QPC model. The relativistic
two-particle states have been thoroughly discussed in the textbooks and
in many papers, such as in
Ref.\cite{macfarlane1963,McKerrell1964,Fuda:2012xd}.

\noindent\emph{Single-particle state:}

Since the transformation between different inertial frames will be
frequently used in this paper, we first define a general
canonical Lorentz boost $l_c(p)$, symbolized by the four-momentum
$p=(p^0,\mathbf p)$. If a four-momentum $q=(q^0,\mathbf q)$ in
an original inertial frame is boosted to an inertial frame moving at
the relative velocity $\mathbf v=-\frac{\mathbf p}{p^0}$ with respect
to the original one, it becomes \bqa
l_c(p)q=l_c(p)\left(\begin{array}{c}
                                q^0 \\
                                \mathbf{q}
                              \end{array}
\right)=\left(\begin{array}{c}
                                \frac{p^0q^0+\mathbf{p}\cdot\mathbf{q}}{W} \\
                                \mathbf{q}+\frac{\mathbf{p}}{W}(q^0+\frac{\mathbf{p}\cdot\mathbf{q}}{p^0+W})
                              \end{array}
\right),
\ \  W=(p.p)^{1/2}.
\eqa

A general single-particle state with its mass $\mu$, momentum $\mathbf
p$, spin $s$, and
third component of spin $m$ is defined by transforming it from its rest
frame to the momentum $\mathbf p$ by a canonical Lorentz boost $l_c(p)$
as
\bqa
|\mathbf p; sm\rangle=U(l_c(p))|\mathbf 0;sm\rangle\sqrt{\frac{\mu}{\varepsilon(\mathbf p)}},
\eqa
where $p=(\varepsilon(\mathbf p),\mathbf p)$ denotes its on-shell
four-momentum, with the energy $\varepsilon(\mathbf p)=(\mathbf
p^2+\mu^2)^{1/2}$. $U(l_c(p))$ represents a unitary operator
representation of the Lorentz boost, and the factor is to ensure the normalization to be
\bqa
\langle \mathbf p,sm|\mathbf p',s'm'\rangle=\delta^{(3)}(\mathbf p-\mathbf p')\delta_{ss'}\delta_{mm'}.
\eqa
This kind of definition is convenient in presenting the coupling form
factor in the relativistic QPC model~\cite{Fuda:2012xd}, which is
required here in the relativistic Friedrichs-Lee model to describe the
coupling between the discrete state and the continuum states.

In the rest frame of a particle, the transformation of the
particle state vector under a spatial rotation $R$ is expressed as
\bqa
U(R)|\mathbf{0},sm\rangle=\sum_{m'}|\mathbf 0, sm'\rangle\mathscr{D}^s_{m'm}(R),
\eqa
where $\mathscr{D}^s_{m'm}(R)$ is the standard matrix representation
of the rotation $R$.
Then the transformation of a general single-particle state with
three-momentum $\mathbf p$ under a general Lorentz
transformation $a$,  will be
\cite{martin}
\bqa
U(a)|\mathbf p, sm\rangle=\sum_{m'}|\mathbf p',sm'\rangle\mathscr{D}^s_{m'm}[r_c(a,p)]\sqrt{\frac{\epsilon(\mathbf
p')}{\epsilon(\mathbf p)}},
\eqa
where $r_c(a,p)=l_c^{-1}(ap)al_c(p)$ is the well-known Wigner rotation.

\noindent\emph{Two-particle state:}

Based on the definition of the canonical single-particle state, one
could obtain the representation of a two-particle
state~\cite{macfarlane1963,McKerrell1964,Fuda:2012xd}, by first
defining it in the c.m. frame of two-particle system and then
boosting it to a general frame. To make it clear, we first define the
momenta in this two frames.

 Consider two particles 1 and 2 with their masses $\mu_1, \mu_2$,
spins $s_1, s_2$, and third components $m_1,m_2$ respectively. In the c.m. frame of a two-particle system, the
four-momenta of two particles are respectively
\bqa
k_1\equiv(k_1^0,\mathbf{k}_1)=(\varepsilon_1(\mathbf k),\mathbf k),\ k_2\equiv(k_2^0,\mathbf{k}_2)=(\varepsilon_2(-\mathbf k),-\mathbf k),
\eqa
where $\mathbf k$ is the three-momentum of particle 1 in the c.m.
frame of the two-particle system, and $\varepsilon_i(\mathbf
k)=(\mathbf k^2+\mu_i^2)^{1/2}$. In a general frame, the four-momenta of two particles are respectively
\bqa
p_1\equiv(p_1^0,\mathbf{p}_1)=(\varepsilon_1(\mathbf p_1),\mathbf
p_1),\quad p_2\equiv(p_2^0,\mathbf{p}_2)=(\varepsilon_2(\mathbf p_2),\mathbf p_2).
\eqa
The relation between the two sets of momenta is
\bqa
p_1=l_c(p)k_1,\quad p_2=l_c(p)k_2,
\eqa
where $p\equiv(p^0,\mathbf{p})=p_1+p_2=(\varepsilon_1(\mathbf
p_1)+\varepsilon_2(\mathbf p_2),\mathbf p_1+\mathbf p_2)$.

A general two-particle state could be defined by boosting the
two-particle state in the c.m frame by $l_c(p)$
\bqa
&&|\mathbf p\mathbf k,s_1s_2 m_1m_2\rangle=U(l_c(p))|\mathbf k,s_1 m_1\rangle\otimes|-\mathbf k,s_2m_2\rangle[W(\mathbf k)/E(\mathbf p,\mathbf k)]^{1/2},\nonumber\\
&&W(\mathbf k)=\varepsilon_1(\mathbf k)+\varepsilon_2(-\mathbf k),\ E(\mathbf{p},\mathbf{k})=(\mathbf{p}^2+W(\mathbf k)^2)^{1/2},
\eqa
where the factor is also to ensure the normalization of the states as
\bqa
\langle \mathbf p\mathbf k,s_1s_2 m_1 m_2|\mathbf p'\mathbf
k',s_1's_2' m_1' m_2'\rangle=\delta^{(3)}(\mathbf p-\mathbf
p')\delta^{(3)}(\mathbf k-\mathbf
k')\delta_{s_1s_1'}\delta_{s_2s_2'}\delta_{m_1m_1'}\delta_{m_2m_2'}.
\eqa
According to the Lorentz transformation properties of the
single-particle states and the standard derivation, one could find
that the
two-particle state mentioned above could be expressed as the
combination of the  direct
product of two single-particle states multiplied with extra factors of
the matrix representations for the Wigner rotations of the two particles
\bqa\label{eq:twoparticlestate}
&&|\mathbf p\mathbf k,s_1s_2 m_1 m_2\rangle=\sum_{m'_1 m'_2}|\mathbf
p_1,s_1 m_1'\rangle\otimes|\mathbf p_2,s_2
m'_2\rangle\mathscr{D}^{s_1}_{m_1'm_1}[r_c(l_c(p),k_1)]\mathscr{D}^{s_2}_{m_2'm_2}[r_c(l_c(p),k_2)](\frac{\varepsilon_1(\mathbf
p_1)}{\varepsilon_1(\mathbf k)}\frac{\varepsilon_2(\mathbf
p_2)}{\varepsilon_2(-\mathbf k)}\frac{W(\mathbf k)}{E(\mathbf
p,\mathbf k)})^{1/2}\,.\nonumber\\
\eqa
Similarly, one can do the partial wave decomposition and couple the
orbital angular momentum $l$ and total spin $s$ to be the total
angular momentum $j$
in the c.m. frame, and then boost to the general frame to obtain
the two-particle state in the angular momentum representation
\bqa
|\mathbf{p}klsjm\rangle&=&\sum_{m_lm_s}\sum_{m_1m_2}\int
d\Omega_{\mathbf{k}}Y_l^{m_l}(\mathbf{\hat{k}})|\mathbf{p}\mathbf{k},s_1s_2m_1m_2\rangle\langle
s_1s_2m_1m_2|sm_s\rangle\langle lsm_lm_s|jm\rangle,
\\
&=&\sum_{m_lm_s}\sum_{m_1m_2}\sum_{m'_1 m'_2}
\int d\Omega_{\mathbf{k}}|\mathbf p_1,s_1 m_1'\rangle\otimes|\mathbf
p_2,s_2 m'_2\rangle Y_l^{m_l}(\mathbf{\hat{k}})\langle
s_1s_2m_1m_2|sm_s\rangle\langle lsm_lm_s|jm\rangle \nonumber
\\
&&\times\mathscr{D}^{s_1}_{m_1'm_1}[r_c(l_c(p),k_1)]\mathscr{D}^{s_2}_{m_2'm_2}[r_c(l_c(p),k_2)](\frac{\varepsilon_1(\mathbf
p_1)}{\varepsilon_1(\mathbf k)}\frac{\varepsilon_2(\mathbf
p_2)}{\varepsilon_2(-\mathbf k)}\frac{W(\mathbf k)}{E(\mathbf
p,\mathbf k)})^{1/2}.
\eqa
with the normalization being
\bqa
\langle \mathbf{p}klsjm|\mathbf{p}'k'l's'j'm'\rangle=\delta^{(3)}(\mathbf p-\mathbf p')\frac{\delta( k-k')}{k^2}\delta_{ll'}\delta_{ss'}\delta_{jj'}\delta_{mm'},
\label{eq:normalization-2p}
\eqa
where $k$ is the magnitude of the three-momentum $\mathbf k$ and
$\mathbf{\hat{k}}$ is direction of $\mathbf k$.  A rigorous and
detailed construction of this state in a common normalization
convention could be found in Ref.~\cite{macfarlane1963,McKerrell1964}.

\subsection{The relativistic Friedrichs-Lee model}\label{relativisticFM}

The Hamiltonian for the non-relativistic coupling of a discrete state and the continuum
state in three dimensional space
can be expressed in a Friedrichs
model~\cite{Xiao:2016mon,Xiao:2016wbs}  after the partial wave
decomposition to the angular momentum representation, where
the final solution contains a non-relativistic dispersion relation,
and so
does the Lee model~\cite{Lee:1954iq}. To extend the model to a
relativistic formalism, we can express the coupling in the creation and annihilation
operators for the single-particle state and for a bilocal state
representing a two-particle state using
the method developed by Antoniou et al. in
ref.~\cite{Antoniou:1998JMP}, where the coupling between a local
Klein-Gordan field with a fixed mass and a bilocal Klein-Gordan field
with a continuum spectrum is considered.
Here, we are going to
consider a bare meson coupled to a meson pair, so the bare meson
could be represented by a single-particle state and the
meson pair could be expressed in the total angular momentum
representation $|\mathbf{p}klsjm\rangle$ and be effectively  mimicked
by a bilocal field but with  more inner degrees of freedom.

Since the normalization of a single-particle state studied here is
$\langle \mathbf{p} jm|\mathbf{p}'
j'm'\rangle=\delta^{(3)}(\mathbf{p}-\mathbf{p}')\delta_{jj'}\delta_{mm'}$,
we can define the creation operator of a single-particle as
\bqa
|\mathbf{p} jm\rangle=a^\dag_{\mathbf pjm}|0\rangle.
\eqa
The commutation relation of the annihilation and creation operators of
the single particle is
\bqa
[a_{\mathbf pjm},a^\dag_{\mathbf p'j'm'}]=\delta^{(3)}(\mathbf{p}-\mathbf{p}')\delta_{jj'}\delta_{mm'}.
\eqa
On the other side, the normalization of two-particle states in the angular momentum representation $|\mathbf{p}klsjm\rangle$ is
\bqa
\langle\mathbf{p}klsjm|\mathbf{p}'k'l's'j'm'\rangle=\delta^{(3)}(\mathbf{p}-\mathbf{p}')\frac{\delta(k-k')}{k^2}\delta_{ll'}\delta_{ss'}\delta_{jj'}\delta_{mm'},
\eqa
so, we can also define the creation operator of a bilocal field,
representing a two-particle state,
$B^\dag_{\mathbf{p}klsjm}$, as
\bqa
|\mathbf{p}klsjm\rangle=B^\dag_{\mathbf{p}klsjm}|0\rangle,
\eqa
with the commutation relation
\bqa
[B_{\mathbf{p}klsjm},B^\dag_{\mathbf{p}'k'l's'j'm'}]=\delta^{(3)}(\mathbf{p}-\mathbf{p}')\frac{\delta(k-k')}{k^2}\delta_{ll'}\delta_{ss'}\delta_{jj'}\delta_{mm'}.
\eqa

To simplify the representation of the formula, we define
$a(\mathbf{p})\equiv a_{\mathbf pjm}$ and $B({\mathbf{p},k})\equiv
B_{\mathbf{p}klsjm}$ with only the variables $\mathbf{p}$ and $k$ kept
and the other variables $lsjm$ omitted in the derivation procedure and
restored finally, since $jm$ should be conserved in the strong
interaction between the discrete state and the continuum state and
$ls$ will symbolize the coupling form factors involved.

If we introduce an interaction between the single-particle state and
the two-particle state, then the full Hamiltonian at $t=0$ or in
Schr\"odinger picture could be expressed
as
\bqa
P_0&=&\int
\mathrm{d}^3\mathbf{p}\omega(\mathbf{p})a^\dag(\mathbf{p})a(\mathbf{p})+\int
\mathrm{d}^3\mathbf{p}k^2\mathrm{d}k
E(\mathbf{p},\mathbf{k})B^\dag({\mathbf{p},k})B({\mathbf{p},k})\nonumber\\
&&+\int \mathrm{d}^3\mathbf{p}k^2\mathrm{d}k
\alpha(k)(a(\mathbf{p})+a^\dag({-\mathbf{p}}))(B^\dag({\mathbf{p},k})+B({-\mathbf{p},k})),
\eqa
where the energy of the single-particle state is
$\omega(\mathbf{p})=(\mathbf{p}^2+\omega_0^2)^{1/2}$ and the total energy
of the two-particle state is
$E(\mathbf{p},\mathbf{k})=(\mathbf{p}^2+W(\mathbf{k})^2)^{1/2}$ with
the c.m. energy defined by
$W(\mathbf{k})=\varepsilon_1(\mathbf{k})+\varepsilon_2(-\mathbf{k})$.
The coupling form factor,  $\alpha(k)$, representing the interaction
between the single-particle state and the two-particle state, in
principle depends on $k$ and the sum over $l, s$ quantum
numbers in the expression should be understood.

It is clearer and convenient to change the variable $k$, the
magnitude of the relative momentum of two particles, to $E$, the total
energy of the two-particle states, in accordance with the ordinary
Friedrichs model. Using the relation
\bqa
\frac{\mathrm{d}k}{\mathrm{d}E}=\frac{Ep_1^0p_2^0}{W^2k},
\eqa
one can obtain the relations
\bqa
k^2\mathrm{d}k=\frac{kEp_1^0p_2^0}{W^2}\mathrm{d}E,\ \ \frac{\delta(k-k')}{k^2}=\frac{W^2}{kEp_1^0p_2^0}\delta(E-E'),
\eqa
where $W=W(\mathbf k)$ and $E=E(\mathbf{p},\mathbf{k})$.
One can define $\beta(E)=\frac{kEp_1^0p_2^0}{W^2}$ to simplify the
following formula. Then the Hamiltonian could be rewritten as
\bqa
P_0&=&\int \mathrm{d}^3\mathbf{p}\int_{M_{th}}\mathrm{d}E\beta(E) EB^\dag(E,\mathbf{p})B(E,\mathbf{p})+\int \mathrm{d}^3\mathbf{p}\omega(\mathbf{p})a^\dag(\mathbf{p})a(\mathbf{p})\nonumber\\
&&+\int \mathrm{d}^3\mathbf{p}\int_{M_{th}}\mathrm{d}E\beta(E) \alpha(k(E,\mathbf p))(a(\mathbf{p})+a^\dag(-\mathbf{p}))(B^\dag(E,\mathbf{p})+B(E,-\mathbf{p})),
\eqa
and the 3-momentum operator is
\bqa
\mathbf P&=&\int \mathrm{d}^3\mathbf{p}\int_{M_{th}}\mathrm{d}E\beta(E) \mathbf{p}B^\dag(E,\mathbf{p})B(E,\mathbf{p})+\int \mathrm{d}^3\mathbf{p}\mathbf{p}a^\dag(\mathbf{p})a(\mathbf{p}),
\eqa
where $M_{th}$ is the energy threshold for the continuum.
The commutation relation of two-particle operators is
\bqa
[B(E,\mathbf{p}),B^\dag(E',\mathbf{p}')]=\delta^{(3)}(\mathbf{p}-\mathbf{p}')\beta(E)^{-1}\delta(E-E').
\eqa

In the non-relativistic Friedrichs model, solving the problem is to
find the solutions of generalized eigenfunction with the complex
eigenvalues for the Hamiltonian~\cite{Xiao:2016mon,Zhou:2017txt}.
 Here, in the relativistic case, the eigenvalue problem is equivalent to
finding the solution of
\bqa
[P_\mu,b^\dag(E,\mathbf{p})]=p_\mu b^\dag(E,\mathbf{p}),
\label{eq:eigeneq}
\eqa
with the creation operator $b^\dag(E,\mathbf{p})$ being written as the
linear superposition of
$B^\dag(E,\mathbf{p})$,$B(E,-\mathbf{p})$,$a^\dag(\mathbf{p})$, and
$a(-\mathbf{p})$ as
\bqa
b^\dag(E,\mathbf{p})&=&\int \beta(E')
\mathrm{d}E'[T(E,E',\mathbf{p})B^\dag(E',\mathbf{p})+R(E,E',\mathbf{p})B(E',-\mathbf{p})]\nonumber\\
&&+t(E,\mathbf{p})a^\dag(\mathbf{p})+r(E,\mathbf{p})a(-\mathbf{p}).
\eqa

By a direct calculation of the commutation relation in
Eq. (\ref{eq:eigeneq}) and comparing
the coefficient of each operator, one finds the relations
\bqa
(E+\omega(\mathbf{p}))r(E,\mathbf{p})=(E-\omega(\mathbf{p}))t(E,\mathbf{p}),\quad
\ (E+E')R(E,E',\mathbf{p})=(E-E')T(E,E',\mathbf{p}).
\eqa
After eliminating $R(E,E',\mathbf{p})$ and $r(E,\mathbf{p})$, one can obtain
\bqa
&(E'-E)T(E,E',\mathbf{p})+\alpha(k)\frac{2\omega(\mathbf{p})}{E+\omega(\mathbf{p})} t(E,\mathbf{p})=\gamma(E)(E'-E)\delta(E'-E),\\
&\int \mathrm{d}E'\beta(E')\alpha(k(E',\mathbf p))\frac{2E'}{E+E'}T(E,E',\mathbf{p})=(E-\omega(\mathbf{p}))t(E,\mathbf{p}),
\eqa
and
\bqa
T(E,E',\mathbf{p})=\gamma(E)\delta(E'-E)-\frac{2\omega(\mathbf{p})\alpha(k(E',\mathbf p))}{(E'-E)(E+\omega(\mathbf{p}))}t(E,\mathbf{p}).
\eqa
Substituting it back into the equation above, one obtains
\bqa
t(E,\mathbf{p})=\frac{\beta(E)\alpha(k(E,\mathbf
p))\gamma(E)(E+\omega(\mathbf{p}))}{\eta_\pm(E,\mathbf{p})},
\eqa
where $\eta_\pm(E,\mathbf{p})$ is expressed as
\bqa
\eta_\pm(E,\mathbf{p})=E^2-\omega(\mathbf{p})^2-\int
\mathrm{d}E'^2[\frac{2\omega(\mathbf{p})\beta(E')\alpha(k(E',\mathbf
p))^2}{(E^2-E'^2\pm i0)}],
\eqa
which appears in the denominator of all the coefficients functions.
We have introduced the $i0$ in the integral to make the integral
well-defined and the $+$($-$) sign will correspond to the
in(out)-state solution.
Similar to the $\eta_\pm(x)$ in the non-relativistic Friedrichs model~\cite{Xiao:2016mon,Zhou:2017txt},
$\eta(E,\mathbf{p})$ is just
the inverse of the resolvent function, which has a right hand cut
starting from threshold energy squared for the two-particle continuum.
In the c.m. frame, $\mathbf{p}=\mathbf{0}$, the variable is changed to
the invariant mass $W$, and the $\eta_\pm(W)$ function reads
\bqa\label{DR:W}
\eta_\pm(W)=W^2-\omega_0^2-\int_{s_{th}} \mathrm{d}W'^2
\frac{\rho(W')}{W^2-W'^2\pm i0}\,,
\label{eq:inverse-resolvent}
\eqa
where $s_{th}=(\mu_1+\mu_2)^2$ and the spectral function
$\rho(W)=2\omega_0\beta(W)\alpha(k)^2=2\omega_0\frac{k\varepsilon_1\varepsilon_2}{W}\alpha(k)^2$
in which the coupling form factor $\alpha(k)$ could be obtained using
some model as we will show in the next subsection. In principle, the coupling form factor should include the interaction of the single-particle state and the two-particle state with different $L$ and $S$, the relative angular momentum and the total spin quantum numbers of the two particles, thus $\alpha(k)^2=\sum_{LS}\alpha_{LS}(k)^2$.   With the variable changed from $W$ to $s$, Eq.~(\ref{DR:W}) could be expressed as
\bqa\label{DR:s}
\eta_\pm(s)=s-\omega_0^2-\int_{s_{th}} ds' \frac{\rho(s')}{s-s'\pm i0},
\eqa
which is Lorentz invariant and just similar to the relativistic dispersion relation.
The main difference from the non-relativistic case is that the relation
is in terms of  the energy squared $s$ instead of the energy $E$.

Thus, we have the in-state creation operator:
\begin{align}
b_{in}^\dagger(E,\mathbf p)=&B^\dagger(E,\mathbf
p)-\frac{2\omega(\mathbf p)\alpha(
k(E,\mathbf p))}{\eta_+(E,\mathbf p)}\bigg[\int_{M_{th}} dE'\beta(E')\alpha(k(E',\mathbf p))\Big(\frac{B^{\dagger}(E',\mathbf p)}{(E'-E-i0)}
-\frac{B(E',-\mathbf p)}{(E'+E+i0)}\Big)
\nonumber
\\&-\frac1{2\omega(\mathbf p)}
\Big((\omega(\mathbf p)+E)a^\dagger(\mathbf p)
-(\omega(\mathbf p)-E)a(-\mathbf p)\Big)\bigg]\,,
\end{align}
which satisfies $[b_{in}(E,\mathbf p),b_{in}^\dagger(E',\mathbf
p')]=\beta^{-1}(E)\delta(E-E')\delta^{(3)}(\mathbf p-\mathbf p')$, and the
normalization $ \gamma(E)=1/\beta(E)$ is determined by this commutation
relation. The out-state
creation operator is similar with all the signs before $i0$ reversed
and the subscript of $\eta$ is also reversed. The vacuum
$|\Omega\rangle$ is also
different from the free cases, and the in-states and out-states are
generated by the creation operators acting on this exact vacuum.
The $S$-matrix of one continuum state can also be obtained by inner product of the in-states
and the out-states,
\begin{align}
S(E,\mathbf p;E',\mathbf p')=\delta^{(3)}(\mathbf p-\mathbf p')\delta(E-E')\Big(1-2\pi
i\frac{\rho(s)}{\eta_+(s)}\Big)\,.
\end{align}

The discrete states are the solution to the $\eta(s)=0$ where
$\eta(s)$ is the analytically continued $\eta_{\pm}$ to
the complex $s$-plane with $\eta_+$ and $\eta_-$ on the upper and lower
rim of the unitarity cut. The discrete states
include the bound states on the real axis of the first sheet, virtual
states on the real axis of the second sheet, and resonances on the
second Riemann sheet of the complex $s$-plane. The creation operators
for the bound states can also be solved to be
\begin{align}
b^\dagger(E_0,\mathbf p)=&N\bigg[\frac{(\omega(\mathbf
p)+E_0)}{\sqrt{2\omega(\mathbf p)}}a^\dagger(\mathbf p)-\frac{(\omega(\mathbf
p)-E_0)}{\sqrt{2\omega(\mathbf p)}}a(-\mathbf p)
\nonumber
\\&-\sqrt{2\omega(\mathbf p)}\int_{M_{th}}
dE'\beta(E') \Big[\frac{\alpha(k(E',\mathbf p))}{E'-E_0} B^\dagger
(E',\mathbf p) -\frac{\alpha(k(E',\mathbf
p))}{E'+E_0} B(E',-\mathbf p)\Big]\bigg],
\end{align}
where the normalization is chosen to be
$N=\frac 1{\sqrt{2E_0}}\Big[1+2\omega(\mathbf p)\int_{M_{th}} dE'
\beta(E')\frac{2E'|\alpha(k(E',\mathbf
p))|^2}{(E'+E_0)^2(E'-E_0)^2}\Big]^{-1/2}$ such that the commutation
relation is
$[b(\mathbf p),b^\dagger(\mathbf p')]=\delta^{(3)}(\mathbf p-\mathbf
p')$. For resonances and virtual states, similar operators can be
found, but since their positions are on the second sheet, the integral
contour in the definition of the operator should be deformed and the
commutator may not be well-defined.

In general, since there is only one continuum state here, i.e. one
unitarity cut,  every
bare discrete state will generate two poles, either becoming a pair of
resonance poles on the second Riemann sheet or remaining on the real axis
being virtual or bound state poles. When the coupling is turned down,
these poles will move back to the bare position of the discrete state.
There could also be dynamically generated poles which does not
move to the bare states and normally will run towards the
singularities of the form factor when the coupling is switched
off~\cite{Xiao:2016mon,Xiao:2016wbs}.
These are dynamically generated by the interaction between the
discrete state and the continuum.

This model can easily be generalized to include more discrete bare
states and continuum bare states. With more discrete bare states, the $\eta$
function will become a matrix whose dimension is equal to the
number of the discrete bare states. With more continuum bare states, more
dispersion integrals will be added in the $\eta$ functions, with each
integral corresponding to a continuum threshold. The number of the
continuum solutions and the dimension of the $S$-matrix is the same as
the number of the bare continuum states. If some continuum states come
with the same threshold, such as those states with only different
isospin but with a degenerate mass when isospin breaking is ignored, the
dispersion integrals for them combine into one, and the $\eta$ function
is still similar to the one for a single continuum. The Riemann sheets is
doubled when a new threshold for the continuum is added. The discrete
state solutions are then the zero points for the determinant of the
$\eta$ matrix. For discrete states originated from the bare discrete
states, the number is also doubled when the Riemann sheets is doubled.
Also, there could be dynamically generated poles which are also doubled
when a new continuum threshold is added. All the poles with the same
origin are called shadow poles~\cite{Eden:1964zz}. In present paper,
for simplicity, we will confine ourselves to the cases with a single
continuum threshold and one discrete state.

\subsection{Coupling form factor from a relativistic QPC model}\label{relqpc}

Now we are going to study the coupling form factor between the bare
meson and the meson-pair states. In the quark potential model, a
meson state is described as the bound state of a valence quark and a
valence anti-quark. Thus, the interaction between a bare meson state
and a meson-pair continuum state could be described by the QPC
model~\cite{Micu:1968mk}, in which a quark and antiquark pair created
from the vacuum and the one in the original meson separate and regroup
to form new mesons. Fuda has generalized
the QPC model to include the relativistic boost effects of the
quarks between different frames~\cite{Fuda:2012xd}. We rewrite
the relativistic QPC model in a more convenient version and
in a more general case where the mesons can have arbitrary
quantum numbers and unequal quark and antiquark masses.

Based on the definition of two-particle state above,
we could write down a relativistic mock state of the bare meson $A$, with 
three-momentum $\mathbf{p}$, mass eigenvalue $\tilde W$, 
orbital angular momentum $l_A$ of two quarks, total spin
$s_A$ of quarks, total angular momentum $j_A$ and its third
component $m_{j_A}$, as
\bqa
|A(\tilde W,^{2s_A+1}{l_A}_{j_Am_{j_A}})(\mathbf{p})\rangle
=\sum_{m_lm_s}\sum_{m_1m_2}\int
d^3\mathbf{k}\psi_{l_Am_{l_A}}^{A}(\mathbf{k})|\mathbf{p}\mathbf{k}s_1s_2m_1m_2\rangle\langle s_1s_2m_1m_2|s_Am_{s_A}\rangle\langle l_As_Am_{l_A}m_{s_A}|j_Am_{j_A}\rangle,
\nonumber \\
\eqa
where $\psi_{l_Am_{l_A}}^{A}(\mathbf{k})$ is the relative wave
function of quarks in the momentum space in the c.m. frame of the
meson, which
could be obtained by solving the eigenfunction in some potential model
as in Ref.\cite{Godfrey:1985xj}.   The normalization of the wave function is
\bqa
\int d^3\mathbf{k} |\psi_{l_Am_{l_A}}^{A}(\mathbf{k})|^2=1.
\label{eq:normalization-A}
\eqa

Furthermore, if the flavor and color indices of quarks are considered,
according to Eq.~(\ref{eq:twoparticlestate}), the meson mock state could be represented as~\cite{Fuda:2012xd}
\bqa
&&|A(\tilde
W,^{2s_A+1}{l_A}_{{j_Am_{j_A}}})(\mathbf{p})\rangle=\sum_{m_lm_s}\sum_{m_1m_2}\sum_{m'_1
m'_2}\int d^3\mathbf{k}\psi_{l_Am_{l_A}}^{A}(\mathbf{k})|\mathbf
p_1,s_1 m_1'\rangle\otimes|\mathbf p_2,s_2
m'_2\rangle\phi_A^{12}\omega_A^{12}\nonumber\\
&&\times\mathscr{D}^{s_1}_{m_1'm_1}[r_c(l_c(p),k_1)]\mathscr{D}^{s_2}_{m_2'm_2}[r_c(l_c(p),k_2)]\langle
s_1s_2m_1m_2|s_Am_{s_A}\rangle\langle
l_As_Am_{l_A}m_{s_A}|j_Am_{j_A}\rangle (\frac{\varepsilon_1(\mathbf
p_1)}{\varepsilon_1(\mathbf k)}\frac{\varepsilon_2(\mathbf
p_2)}{\varepsilon_2(-\mathbf k)}\frac{W_{12}(\mathbf
k)}{E_{12}(\mathbf p,\mathbf k)})^{1/2}.\nonumber\\
\eqa
The subscript 1 and 2 refer to the quark and antiquark in meson $A$
respectively, $W_{12}(\mathbf
k)=\varepsilon_1(\mathbf k)+\varepsilon_2(-\mathbf k)$, and $E_{12}(\mathbf p,\mathbf k)=\sqrt{W_{12}(\mathbf
k)^2+\mathbf p^2}$, while $\phi_A^{12}$ denotes the flavor wave function and
$\omega_A^{12}$ the color wave function of the meson.

In the relativistic QPC model~\cite{Fuda:2012xd}, an instant interaction Hamiltonian
\bqa
H_I=\gamma\int \mathrm{d}^3x\bar\psi(x)\psi(x),\ \ t=0,
\eqa
is assumed, where $\psi(x)$ is a Dirac field operator at $x$. $\gamma$
is the strength parameter representing the quark pair production from
the vacuum. Then, the transition operator could be derived and written down as
\bqa
&&T=-\sqrt{8\pi}\gamma\int\frac{ d^3\mathbf p_3d^3\mathbf
p_4}{\sqrt{\varepsilon_3(\mathbf p_3)\varepsilon_4(\mathbf
p_4)}}\delta^{(3)}(\mathbf p_3+\mathbf p_4)\sum_m\sum_{m_3m_4}\langle 1,m,1,-m|0,0\rangle\nonumber\\
&&\times\mathscr{Y}_1^m(\frac{\mathbf p_3-\mathbf p_4}{2})\langle 1/2,m_3,1/2,m_4|1,-m\rangle \phi_0^{34}\omega_0^{34}b^\dag_{m_3}(\mathbf p_3)d^\dag_{m_4}(\mathbf p_4),
\eqa
where the subscript 3 and 4 refer to the
quark and the anti-quark produced from the vacuum respectively.
$\phi_0^{34}$ and $\omega_0^{34}$ are the flavor and color wave
functions of the quark pair from the vacuum.
$\mathscr{Y}_1^m(\frac{\mathbf p_3-\mathbf p_4}{2})$ is the solid
harmonics. $b^\dag_{m_3}$ and $d^\dag_{m_4}$ are the creation operators
of the quark and the anti-quark.

\begin{figure}
  \centering
  \includegraphics[width=48mm]{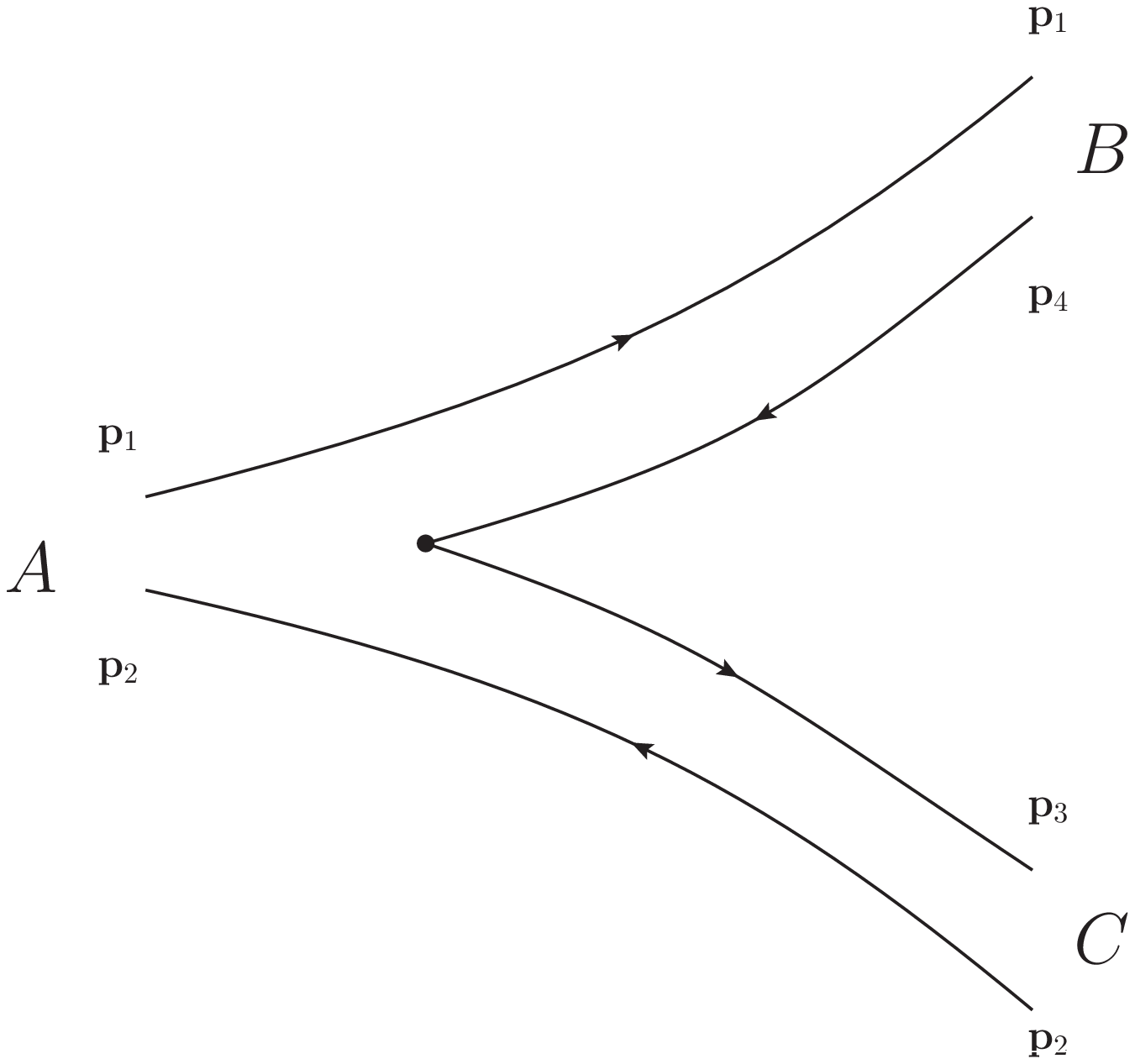}\hspace{5mm}
  \includegraphics[width=52mm]{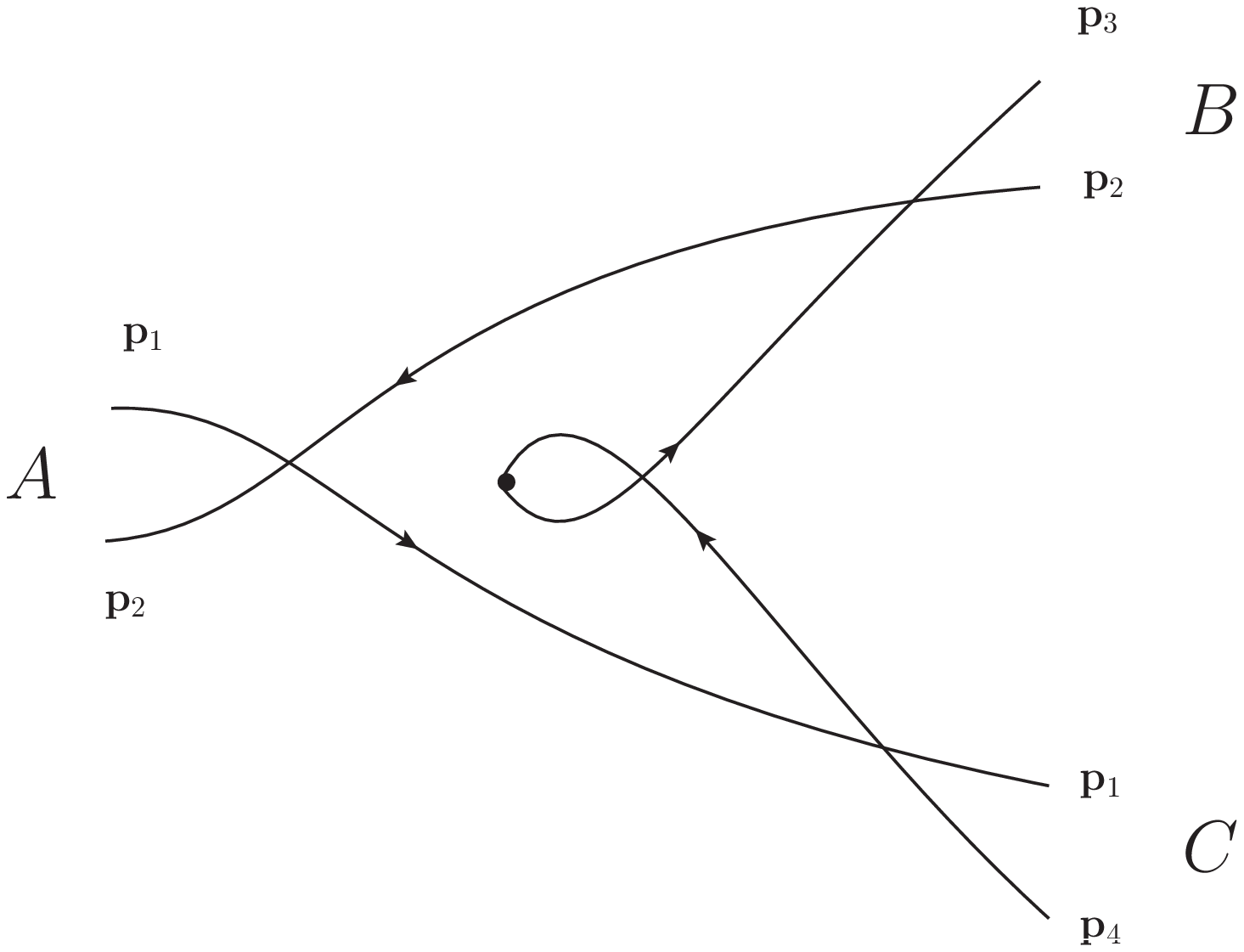}\\
  \caption{Two kinds of diagrams which could happen in the quark pair creation model. The arrows on the quark lines only represent the directions of fermion lines. Usually only one of the diagrams will contribute to the amplitude, but both of them will have contributions for the $f_0$ meson discussed here. }\label{qpc}
\end{figure}

If we define the $S$-matrix of  $A\to BC$ process as
\bqa
S=I-2\pi i \delta(E_A-E_B-E_C)\delta^{(3)}(\mathbf{P}_A-\mathbf{P}_B-\mathbf{P}_C)\mathscr{M}^{m_{j_A}m_{j_B}m_{j_C}},
\eqa
then the $A\to BC$ amplitude in the c.m. frame of meson $A$ could be expressed as
\bqa\label{eq:helicityM}
&&\mathscr{M}^{m_{j_A}m_{j_B}m_{j_C}}(\mathbf{q})=\sum_{\tiny m_{l_B}m_{s_B}m_{l_C}m_{s_C}m_{l_A}m_{s_A}m}\langle l_As_Am_{l_A}m_{s_A}|j_Am_{j_A}\rangle\langle  l_Bs_Bm_{l_B}m_{s_B}|j_Bm_{j_B}\rangle\nonumber\\
 &&\times \langle l_Cs_Cm_{l_C}m_{s_C}|j_Cm_{j_C}\rangle\langle 1,m,1,-m|0,0\rangle\nonumber\\
 &&\times\{\langle\phi_B^{14}\phi_C^{32}|\phi_0^{34}\phi_A^{12}\rangle\int \mathrm{d}^3\mathbf{k} \frac{(-\sqrt{8\pi}\gamma/3)}{\varepsilon_3(\mathbf p_3)}\psi^{B*}_{l_Bm_{l_B}} (\mathbf{k})\psi^{C*}_{l_Cm_{l_C}}(\mathbf{k}')\psi^{A}_{l_Am_{l_A}}(\mathbf{p}_1)\mathscr{Y}_1^m(\mathbf{p}_3)\nonumber\\
&&\times\sum_{\tiny\begin{array}{c}
                m_1m_4m_3m_2 \\
               m'_1m'_4m'_3m'_2
              \end{array}
}\langle s_1s_2m_1'm_2'|s_Am_{s_A}\rangle\langle s_1s_4m_1m_4|s_Bm_{s_B}\rangle\langle s_3s_2m_3m_2|s_Cm_{s_C}\rangle\langle s_3s_4m_3'm_4'|1,-m\rangle\nonumber\\
&&\times\mathscr{D}^{(1/2)*}_{m_1'm_1}[r_c(l_c(q_{1}),k_{1})]\mathscr{D}^{(1/2)*}_{m_4'm_4}[r_c(l_c(q_{1}),k_{4})]\mathscr{D}^{(1/2)*}_{m_3'm_3}[r_c(l_c(q_2),k_3)]\mathscr{D}^{(1/2)*}_{m_2'm_2}[r_c(l_c(q_2),k_2)]\nonumber\\
&&\times{(\frac{\varepsilon_1(\mathbf p_1)}{\varepsilon_1(\mathbf k)}\frac{\varepsilon_4(\mathbf p_4)}{\varepsilon_4(\mathbf k)}\frac{W_{14}(\mathbf k)}{E_{14}(\mathbf q,\mathbf k)})^{1/2}(\frac{\varepsilon_3(\mathbf k')}{\varepsilon_3(\mathbf p_3)}\frac{\varepsilon_2(\mathbf k')}{\varepsilon_2(\mathbf p_2)}\frac{E_{32}(-\mathbf q,\mathbf k')}{W_{32}(\mathbf k')})^{1/2}}\nonumber\\
&&+\langle\phi_B^{32}\phi_C^{14}|\phi_0^{34}\phi_A^{12}\rangle\int \mathrm{d}^3\mathbf{k}' \frac{(-\sqrt{8\pi}\gamma/3)}{\varepsilon_3(\mathbf p_3)}\psi^{B*}_{l_Bm_{l_B}} (\mathbf{k}')\psi^{C*}_{l_Cm_{l_C}}(\mathbf{k})\psi^{A}_{l_Am_{l_A}}(\mathbf{p}_1)\mathscr{Y}_1^m(\mathbf{p}_3)\nonumber\\
&&\times\sum_{\tiny\begin{array}{c}
                m_1m_4m_3m_2 \\
               m'_1m'_4m'_3m'_2
              \end{array}
}\langle s_1s_2m_1'm_2'|s_Am_{s_A}\rangle\langle s_3s_2m_3m_2|s_Bm_{s_B}\rangle\langle s_1s_4m_1m_4|s_Cm_{s_C}\rangle\langle s_3s_4m_3'm_4'|1,-m\rangle\nonumber\\
&&\times\mathscr{D}^{(1/2)*}_{m_3'm_3}[r_c(l_c(q_{1}),k_{3})]\mathscr{D}^{(1/2)*}_{m_2'm_2}[r_c(l_c(q_{1}),k_{2})]\mathscr{D}^{(1/2)*}_{m_1'm_1}[r_c(l_c(q_2),k_1)]\mathscr{D}^{(1/2)*}_{m_4'm_4}[r_c(l_c(q_2),k_4)]\nonumber\\
&&\times{(\frac{\varepsilon_3(\mathbf p_3)}{\varepsilon_3(\mathbf k')}\frac{\varepsilon_2(\mathbf p_2)}{\varepsilon_2(\mathbf k')}\frac{W_{32}(\mathbf k')}{E_{32}(\mathbf q,\mathbf k')})^{1/2}(\frac{\varepsilon_1(\mathbf k)}{\varepsilon_1(\mathbf p_1)}\frac{\varepsilon_4(\mathbf k)}{\varepsilon_4(\mathbf p_4)}\frac{E_{14}(-\mathbf q,\mathbf k)}{W_{14}(\mathbf k)})^{1/2}}\}
\eqa
with the two terms in the bracket corresponding to two different
diagrams in Fig.~\ref{qpc}.
The factor $1/3$ comes from the overlap of the color wave functions.
Usually, only one of the diagrams is needed.
$\mathbf{k}$ is the three-momentum of particle $1$ in the
c.m. frame of $14$ system, and $\mathbf{k}'$ is the three-momentum of
particle $3$ in the c.m. frame of $32$ system. If particle $1$ and $2$ are of
the same mass, $\mathbf{k}=\mathbf{k}'$, and the normalization
factors, such as the one in the last line of Eq.~(\ref{eq:helicityM}), will cancel.
Notice that the momenta of the quarks in the integrals are different in the
two cases, for example, $\mathbf p_3=\mathbf p_1-\mathbf q$ for the
first case and $\mathbf p_3=\mathbf p_1+\mathbf q$ for the second one. The case that particle 1
and 2 are not of the same flavor is introduced in the Appendix~\ref{appendixB}.
$\mathbf{p}_1$ is the three-momentum of the quark $1$ in the c.m. frame
of meson $A$, and $\mathbf{q}$ is the three-momentum of meson $B$ in
the c.m. frame of $A$. $q_1$ and $q_2$ represent the four-momenta of meson $B$ and $C$ respectively.

If we choose the direction of meson $B$ along the $z$-direction, the
amplitude with the $BC$ system having relative angular momentum $L$ and the total spin
$S$,  is expressed
as~\cite{Jacob:1959at}
\bqa
\mathscr{M}^{LS}(\mathbf q_z)=\frac{\sqrt{4\pi(2L+1)}}{2j_A+1}\sum_{m_{j_B},m_{j_C}}\langle LS0(m_{j_B}+m_{j_C})|j_A(m_{j_B}+m_{j_C})\rangle\nonumber\\
\langle j_B j_C m_{j_B}m_{j_C}|S(m_{j_B}+m_{j_C})\rangle\mathscr{M}^{(m_{j_A}=m_{j_B}+m_{j_C})m_{j_B}m_{j_C}}(\mathbf q_z),
\eqa
which corresponds to the coupling form factor $\alpha_{LS}(k)$ used in the
relativistic Friedrichs-Lee model as in Eq.~(\ref{DR:s}) in the center
of mass system. Since our
normalization for the particle $A$ is $\langle \mathbf p|\mathbf
p'\rangle=\delta^{(3)}(\mathbf p-\mathbf p')$ and the one for particle $BC$
is Eq.~(\ref{eq:normalization-2p}), the quantity $\sqrt{2\omega(\mathbf
p_A)\beta(E)}\alpha_{LS}(E)$ is Lorentz invariant, which is just the
$\sqrt{\rho(s)}$ in Eq.~(\ref{eq:inverse-resolvent}).

\section{A phenomenological example: low lying  $0^{++}$ scalars with
$I=0$}\label{discussion}

In the QPC model described above, the relative wave
function of quarks in a meson and the bare mass of the meson state could
be obtained by solving the quark potential model. In
the GI model~\cite{Godfrey:1985xj}, the Hamiltonian is
modified to incorporate the relativistic effects as
\bqa
\tilde H=(p^2+m_1^2)^{1/2}+(p^2+m_2^2)^{1/2}+\tilde H^{conf}_{12}+\tilde H^{hyp}_{12}+\tilde H^{so}_{12}.
\eqa
Thus, the eigenfunction of $H$ can be used in the relativistic QPC model to represent
the relative wave
functions for the quarks in the c.m. frame and the mass eigenvalues can be identified
with  the bare
masses of the mesons in a consistent manner.
After
numerically diagonalizing the GI's Hamiltonian with the original GI's parameters
by choosing a large number of the simple harmonic
oscillator~(SHO) bases, one could obtain the eigenvalues and eigen-wavefunctions of all the bare $q\bar q$ bound states.
The lowest isoscalar $(u\bar u+d\bar d)/\sqrt{2}$ bound
state predicted by the GI model is located at about 1.1 GeV. On the
other side, the lightest isoscalar state in the PDG table is the
$f_0(500)$, whose pole position is at about $475^{\pm 75}-i 275^{\pm
75}$ MeV.
As was pointed out in the original GI's
paper~\cite{Godfrey:1985xj}, the meson solutions in
the GI potential model are just the quark-antiquark bound states formed
by considering the interaction potentials between the quark and
antiquark,  while the interactions with their decaying channels are
omitted. In fact, if the coupling to the continuum
states~(decaying channels) is considered, two kinds of  consequences
may happen. The first, which always happens, is the mass shift of the discrete state caused
by the ``renormalization'' effect, and the second,  more
importantly,  is
 the emergence of extra poles as discussed in ref.~\cite{Xiao:2016dsx}.
Thus, the second fact suggests us the possibility that the lightest
isoscalar state could be dynamically generated by the interaction
between the discrete state and the continuum. In principle, the
direct coupling of the continuum to continuum also contributes to the
scattering $S$-matrix. However, since the low energy $\pi\pi$
interaction is almost saturated by a $f_0(500)$ resonance, as long as
this resonance is produced, almost all low energy $\pi\pi$ interaction
is included and the residual continuum-continuum interaction will not
contribute much. We will see that the $f_0(500)$ really will be automatically
generated by the interaction of the lowest isoscalar $(u\bar u+d\bar
d)/\sqrt{2}$ bound state and the continuum. Thus the residual
continuum-continuum interaction would be weak compared to the
seed-continuum interaction and will only renormlize the pole position
a little. Technically, including a most general
continuum-continuum interaction will render the Friedrichs-Lee model
unsolvable. So we will ignore such continuum-continuum interaction in
our discussion.

\begin{figure}
  \centering
  \includegraphics[width=10cm]{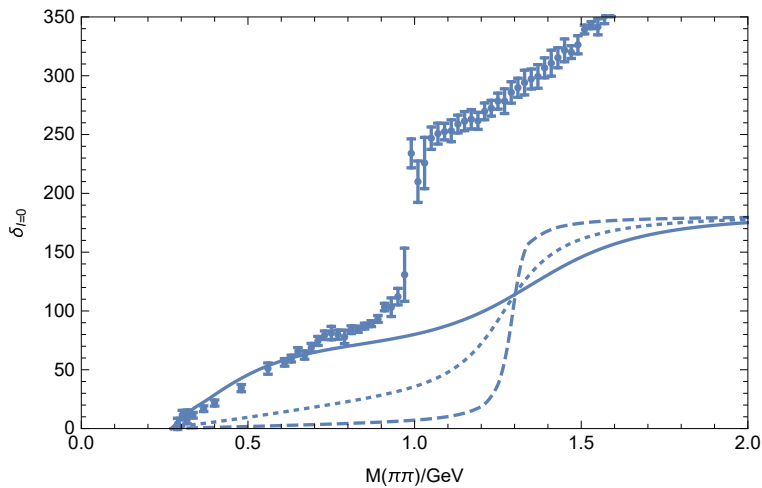}\\
  \caption{The phase shifts caused by the isoscalar $(u\bar u+d\bar d)/\sqrt{2}$ state with $\gamma\simeq 1.4~(dashed), 2.9~(dotted), 4.3~(solid)$GeV, respectively, compared with the experimental data~\cite{Protopopescu:1973sh,Grayer:1974cr,Becker:1978ks}.}\label{phasepipi}
\end{figure}
\begin{figure}
  \centering
  \includegraphics[width=10cm]{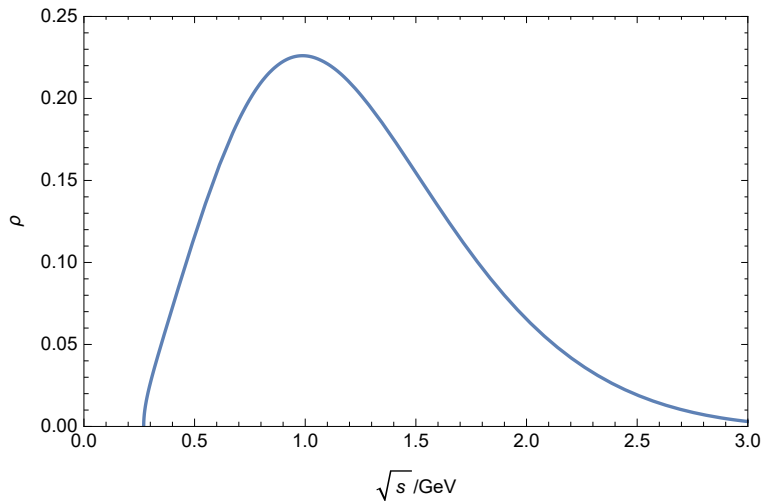}\\
  \caption{The spectral function $\rho(s)$ as a function of $\sqrt s$ for $\pi\pi$
scattering.}\label{fig:rho}
\end{figure}

For simplicity, we consider only single channel cases here, i.e. only one
continuum state.
The lowest isoscalar $(u\bar u+d\bar d)/\sqrt{2}$ bare state is assumed to couple
to the $\pi\pi$ continuum in the QPC model. The wave
functions obtained from the GI model are applied to determine the coupling form factors in the Friedrichs-Lee model.
The $\eta(s)$ function of Eq.~(\ref{DR:s}), being the most important
ingredient of the Friedrichs-Lee model, will serve to provide most
information to be compared with the experiment. When a
continuum state~(decaying channel) is
considered, the $\eta(s)$
function has a unitarity cut starting from the threshold $s_{th}$.
As the  $\eta(s)$
function is continued to the complex $s$-plane on the unphysical
Riemann sheet,
\bqa\label{DR:sII}
\eta^{II}(s)=s-\omega_0^2-\int_{s_{th}} ds' \frac{\rho(s')}{s-s'}-2\pi i\rho(s),
\eqa
the zero points of $\eta^{II}(s)$ function
are just the pole positions of the scattering $S$-matrix.
The spectral function $\rho(s)$ as a function of $\sqrt s$ is shown
in Fig.~\ref{fig:rho}. Secondly, the elastic scattering $S$-matrix is usually parameterized
as $S(s)=e^{2i\delta(s)}$, where $\delta(s)$ denotes the scattering phase
shift. Since the $\eta(s)$ function is just the denominator of
$S(s)$, the scattering phase shift could be represented by the
phase of the $\eta(s)$ function.

To obtain a better description of the pole positions and experiment phase shifts, we slightly change the bare mass of $(u\bar u+d\bar d)/\sqrt{2}$ state to $1.3$ GeV~\cite{Zhou:2020pyo} for the reason mentioned above.
With the $\gamma$ parameter increasing from 0 to a certain value, the
phase shift of isoscalar $\pi\pi$ scattering will exhibit different
behaviors. We only present three cases with $\gamma\simeq 1.4,\
2.9,\ 4.3$GeV, respectively, as shown in Fig.~\ref{phasepipi}. When
$\gamma$ is small, the phase shift looks like a contribution of a
typical narrow resonance or a Breit-Wigner formula, which rises rapidly
to about $180^\circ$ at the vicinity of the mass of the bare state.
When $\gamma$ becomes large, the phase shift will not behave like a
narrow resonance. When $\gamma$ is  about $4.3$GeV, the phase shift in the
lower region will exhibit a mildly rising behavior which could
be identified as a broad $f_0(500)$, as shown in Fig.~\ref{phasepipi}.

Analysis of the pole positions on the complex $s$-plane will help us
understand the behavior.
In fact, two pairs of resonance poles are found on the
unphysical Riemann sheet of the complex $s$-plane. As $\gamma=4.3$GeV,
two zero points extracted from the $\eta^{II}(s)$ function are just
located at about
\bqa
\sqrt{s_{1}}=390-i 255 \mathrm{MeV},\ \  \sqrt{s_{2}}=1349-i 296 \mathrm{MeV}.
\eqa
 The lower pole is just close to the average values of $f_0(500)$ in
the PDG table. It is these $f_0(500)$ and $f_0(1370)$ poles that contribute a smooth
rising of the phase shift, which is the confusing ``red dragon'' about
twenty years ago~\cite{Minkowski:1998mf}. Of course, one could tune the parameters to obtain a better description of the data similar to the
experimentally measured behavior below the $K\bar K$ threshold, which
rises smoothly
and approaches $90^\circ$ at about 850 MeV. However, the more precise
way is to do a combined fit together with the other mesons, taking all the bare
masses and the universal $\gamma$ as the parameters and also including the coupled channel effects,
which is beyond our present work. In
this work, we
only wish to present the general properties of scalar mesons, and the
results here is enough for our purpose.

\begin{figure}
  \centering
  \includegraphics[width=10cm]{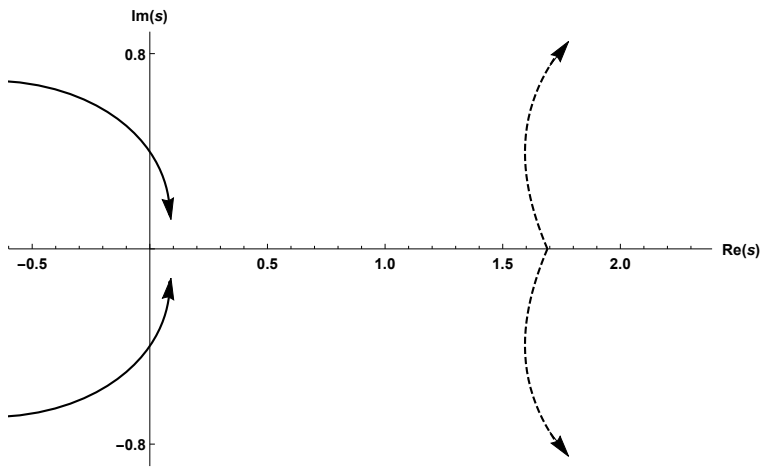}\\
  \caption{Trajectories of the two poles related to the lightest $I=0$
$(u\bar u+d\bar d)/\sqrt{2}$ bare states on the second sheet of the complex $s$-plane.
As $\gamma$ increases from 0 to $4.3$GeV, the bare state will move to the
complex plane and become a broad resonance. Another pole, which
originates from deep in the complex plane, will come to a certain place on the
complex $s$-plane and behave as the $f_0(500)$.
}\label{poletrajsigma}
\end{figure}

The pole trajectories could provide more insights into the nature of
these poles. When $\gamma$ equals 0, which means that the coupling to the
continuum state is not turned on, there is only one pole located at
the bare mass of the discrete state on the real axis. Once $\gamma$
obtains a tiny value, the pole of the discrete state~(referred to as
the ``bare" pole) will move from the real axis to the complex $s$-plane
on the unphysical Riemann sheet and become a resonance. At the
same time, another pair of complex poles come into play with very
large imaginary parts on the complex $s$-plane. These poles does
not exist when $\gamma$ vanishes, so they are dynamically
generated~(referred to as the ``dynamical" pole). As the coupling
strength $\gamma$ increases, the ``bare" poles move away from the real
axis and its imaginary part become larger and larger, while the
``dynamical" ones move close to the real axis with its imaginary part
decreasing, as shown in Fig.~\ref{poletrajsigma}.

The higher pole corresponding to the bare isoscalar
$(u\bar u+d\bar d)/\sqrt{2}$ state might be the $f_0(1370)$.
Since we only consider one continuum here, the pole position may not
be quite precise. It was known
that a mysterious
property of $f_0(1370)$ is that there is no phase shift measured so
that its existence is questioned~\cite{Klempt:2007cp}. From the point
of view here, $f_0(500)/\sigma$ and $f_0(1370)$ appear together. $f_0(500)/\sigma$ is
dynamically generated and $f_0(1370)$ is originated from the bare seed
and they both are very broad. Their contributions to the
$IJ=00$ $\pi\pi$ scattering phase shifts are consistent with the
experiment values in quality up to about 0.9 GeV. In the higher region
above 0.9 GeV, the contributions of $f_0(980)$ and $K\bar K$ threshold
will be important. Taking the $f_0(980)$ into account needs the
formalism of the Friedrichs-Lee model with multiple bare states and
multiple continuous states, which is beyond the scope of this paper.
However, roughly speaking, the difference of phase shifts between the
single-channel approximation here and the experimental values could
just be compensated by  another $180^\circ$ contributed by the
$f_0(980)$. So, it is instructive to look at the total contribution
of $f_0(500)/\sigma$ and $f_0(1370)$ to
the phase shift in this single channel approximation.
  It can be observed from
Fig.~\ref{phasepipi} that the two poles together contribute a rough
$180^\circ$ phase shift, a very mild phase shift from the $\pi\pi$
threshold to about 1.5 GeV.
In fact, this is a rather general property and can be understood as
follows.  Since the $T$ matrix is proportional to the spectral
function
which
 goes to zero as $s\to \infty$, $T$ also goes to zero in this
limit.  For single channel scatterings, $T\propto \sin\delta
e^{i\delta}$, thus in
this limit $\delta$ can only take the value of $n\times 180^\circ$, $n\in \mathbb
Z$. In this case, the total phases contributed by the poles
falls between $0$ and $180^\circ$ and goes monotonically up. Thus the only
limit of the phase shift should be $180^\circ$. This $180^\circ$ phase shift can be
attributed to the two pair of poles, one from the bare state and the
other from the dynamically generated one. Thus, the $\sigma$ and
$f_0(1370)$ together contribute a total phase shift of $180^\circ$.
This
means that they are dynamically related and cannot be treated
as independent.
In general, the above argument is not limited to this scheme.  When single
channel approximation is good, similar argument could be applied to the cases where the denominator of the single
channel $S$-matrix
is just similar to the $\eta$ function here, i.e.  a real
function plus the  dispersion relation
integral where the spectral function goes to zero as $s\to \infty$.

\section{Summary}
In this paper, we proposed a framework to study the hadron spectrum by
generalizing the relativistic Friedrichs-Lee model in a more realistic
scenario and  combining it with the relativistic QPC model in a
consistent way. In the relativistic Friedrichs-Lee model, by
assuming the creation and annihilation operators for a single-particle
bare state and a two-particle bare state and considering the interaction between them, the
exact solution of the creation and annihilation operators for both the single-particle
and the two-particle energy eigenstates could be derived.
Fuda's
relativistic formulation of the QPC model is also generalized to
the  cases with unequal quark-antiquark masses.
The relativistic
exactly-soluble Friedrichs-Lee model combined with the relativistic QPC model and GI's model,
could be used to study the hadron states with light quarks as well
as the ones with heavy quarks in a
relativistically consistent way and in a unified framework.
 This scheme may shed more light on the natures of the light scalar states
in the constituent quark picture~\cite{Zhou:2020pyo}.
As an example, we present that the light $f_0(500)$ and $f_0(1370)$ could
be two poles related to the same bare state, the lightest isoscalar
$(u\bar u+d\bar d)/\sqrt{2}$ state: $f_0(500)$ dynamically generated by the
interaction between the bare state and the $\pi\pi$ continuum, and the
$f_0(1370)$ originated from the bare state.
This scheme might also be helpful in studying the other light meson
states.

\begin{acknowledgments}
Helpful discussions with  Hai-Qing Zhou, Gang Li, and Feng-kun Guo are appreciated. This work is supported by China National Natural
Science Foundation under contract No. 11975075, No. 11575177, and No.11947301. Z.Z is also supported by the Natural Science Foundation of Jiangsu Province of China under contract No. BK20171349.
\end{acknowledgments}
\appendix
\section{Lorentz transformation and kinematics}\label{appendixB}

In this paper, we consider quark ``1'' and antiquark ``2'' in meson $A$
and quark ``3" and antiquark ``4" generated from the vacuum.  Quark
``1" and antiquark ``4" are regrouped to form a meson, so do quark ``3" and
``2". In the case $A(12)\rightarrow B(14)C(32)$,  we define the
four-momenta of quark ``1" and antiquark ``4" in the c.m. frame of
meson $B$ as
$k_1=(\varepsilon_1(\mathbf k),\mathbf k)$,
$k_4=(\varepsilon_4(-\mathbf k),-\mathbf k)$, and the four-momenta of
quark ``3" and ``2" in the c.m. frame of meson $C$ as
$k_3=(\varepsilon_3(\mathbf k'),\mathbf k')$,
$k_2=(\varepsilon_2(-\mathbf k'),-\mathbf k')$. The total four-momenta
of quark and antiquark in the meson mock states $B$ and $C$ are
respectively
\bqa
&&q_1=(\sqrt{(\varepsilon_1(\mathbf k)+\varepsilon_4(-\mathbf k))^2+\mathbf q^2},\mathbf q),\nonumber\\
&&q_2=(\sqrt{(\varepsilon_3(\mathbf k')+\varepsilon_2(-\mathbf k'))^2+(-\mathbf q)^2},-\mathbf q),
\eqa
where $\mathbf q$ is the corresponding total three-momentum in meson $B$.
Then, the Lorentz transformation properties between four-momenta $k_i$ and $p_i$ obey the following relations
\bqa
&&l_c(q_1)k_1=(\varepsilon_1(\mathbf p_1),\mathbf p_1),\nonumber\\
&&l_c(q_1)k_4=(\varepsilon_4(\mathbf p_4),\mathbf p_4),\nonumber\\
&&l_c(q_2)k_3=(\varepsilon_3(\mathbf p_3),\mathbf p_3),\nonumber\\
&&l_c(q_2)k_2=(\varepsilon_2(\mathbf p_2),\mathbf p_2).
\eqa
In the c.m. frame of meson $A$, $\mathbf p_1$ is equal to the relative momentum of quark-antiquark in meson mock state $A$
\bqa
\mathbf p_1=-\mathbf p_2=\mathbf p.
\eqa
The momenta of the quark-antiquark created from the vacuum satisfy $\mathbf p_3=-\mathbf p_4$. Then, $\mathbf{p}_1+\mathbf p_4=\mathbf q$, $\mathbf{p}_3+\mathbf p_2=-\mathbf q$.

The Lorentz transformations of all four quarks are expressed explicitly as
\bqa\label{App4}
&&p_1=l_c(q_1)k_1=l_c(q_1)\left(\begin{array}{c}
                                \varepsilon_1(\mathbf{k}) \\
                                \mathbf{k}
                              \end{array}
\right)=\left(\begin{array}{c}
                                \frac{E_{14}(\mathbf{q},\mathbf{k})\varepsilon_1(\mathbf{k})+\mathbf{p}\cdot\mathbf{k}}{W_{14}(\mathbf{k})} \\
                                \mathbf{k}+\frac{\mathbf{q}}{W_{14}(\mathbf{k})}(\varepsilon_1(\mathbf{k})+\frac{\mathbf{q}\cdot\mathbf{k}}{E_{14}(\mathbf{q},\mathbf{k})+W_{14}(\mathbf{k})})
                              \end{array}
\right),\nonumber\\
&&p_4=l_c(q_1)k_4=l_c(q_1)\left(\begin{array}{c}
                                \varepsilon_4(-\mathbf{k}) \\
                                -\mathbf{k}
                              \end{array}
\right)=\left(\begin{array}{c}
                                \frac{E_{14}(\mathbf{q},-\mathbf{k})\varepsilon_4(-\mathbf{k})-\mathbf{p}\cdot\mathbf{k}}{W_{14}(-\mathbf{k})} \\
                                -\mathbf{k}+\frac{\mathbf{q}}{W_{14}(-\mathbf{k})}(\varepsilon_4(-\mathbf{k})-\frac{\mathbf{q}\cdot\mathbf{k}}{E_{14}(\mathbf{q},-\mathbf{k})+W_{14}(-\mathbf{k})})
                              \end{array}
\right),\nonumber\\
&&p_3=l_c(q_2)k_3=l_c(q_2)\left(\begin{array}{c}
                                \varepsilon_3(\mathbf{k'}) \\
                                \mathbf{k'}
                              \end{array}
\right)=\left(\begin{array}{c}
                                \frac{E_{32}(-\mathbf{q},\mathbf{k'})\varepsilon_3(\mathbf{k'})-\mathbf{p}\cdot\mathbf{k'}}{W_{32}(\mathbf{k'})} \\
                                \mathbf{k'}-\frac{\mathbf{q}}{W_{32}(\mathbf{k'})}(\varepsilon_3(\mathbf{k'})-\frac{\mathbf{q}\cdot\mathbf{k'}}{E_{32}(-\mathbf{q},\mathbf{k'})+W_{32}(\mathbf{k'})})
                              \end{array}
\right),\nonumber\\
&&p_2=l_c(q_2)k_2=l_c(q_2)\left(\begin{array}{c}
                                \varepsilon_2(-\mathbf{k'}) \\
                                -\mathbf{k'}
                              \end{array}
\right)=\left(\begin{array}{c}
                                \frac{E_{32}(-\mathbf{q},-\mathbf{k'})\varepsilon_2(-\mathbf{k'})+\mathbf{p}\cdot\mathbf{k'}}{W_{32}(-\mathbf{k'})} \\
                                -\mathbf{k'}-\frac{\mathbf{q}}{W_{32}(-\mathbf{k'})}(\varepsilon_2(-\mathbf{k'})+\frac{\mathbf{q}\cdot\mathbf{k'}}{E_{32}(-\mathbf{q},-\mathbf{k'})+W_{32}(-\mathbf{k'})})
                              \end{array}
\right).
\eqa

In the equal-mass case, i.e. when quark ``1'' and antiquark ``2'' have
the same mass, one can obtain $\mathbf{k}=\mathbf{k'}$ because $\mathbf{p_1}=-\mathbf{p_2}$ or $\mathbf{p_3}=-\mathbf{p_4}$.

In the unequal-mass case, one should use the inverse relation of the third one in Eq.~(\ref{App4}) to obtain the representation of $\mathbf k'$.
Because $q_2=(q_2^0,-\mathbf q)=(\varepsilon_2(-\mathbf{p_1})+\varepsilon_3(\mathbf{p_1}-\mathbf{q}),-\mathbf{q})$ and
\bqa
&&\left(\begin{array}{c}
                                \varepsilon_3(\mathbf{k'}) \\
                                \mathbf{k'}
                              \end{array}
\right)=l_c(q_2)^{-1}p_3=l_c(q_2)^{-1}\left(\begin{array}{c}
                                \varepsilon_3(\mathbf{p_3}) \\
                                \mathbf{p_3}
                              \end{array}
\right)=l_c(q_2)^{-1}\left(\begin{array}{c}
                                \varepsilon_3(-\mathbf{p_4}) \\
                                -\mathbf{p_4}
                              \end{array}
\right),
\eqa
one could obtain
\bqa
\mathbf k'=-\mathbf p_4+\frac{\mathbf q}{W}\Big(\varepsilon_3(-\mathbf
p_4)-\frac{\mathbf q\cdot\mathbf p_4}{q_2^0+W}\Big),
\eqa
where $ W=\sqrt{q_2\cdot q_{2}}$ and $q_2^0=\varepsilon_2(-\mathbf{p_1})+\varepsilon_3(\mathbf p_1-\mathbf q)$. Thus, $\mathbf k'$ could be expressed as a function of $\mathbf q$ and $\mathbf k$, which could be easily used in the relativistic QPC model.

\bibliographystyle{apsrev4-1}
\bibliography{Ref}

\end{document}